\documentstyle[11pt]{article}
\advance \topmargin by -\headheight
\advance \topmargin by -\headsep
     
\evensidemargin \oddsidemargin
\def\rf#1{(\ref{eq:#1})}
\def\lab#1{\label{eq:#1}}
\def\nonu{\nonumber}
\def\br{\begin{eqnarray}}
\def\er{\end{eqnarray}}
\def\be{\begin{equation}}
\def\ee{\end{equation}}

\def\foot#1{\footnotemark\footnotetext{#1}}
\def\lb{\lbrack}
\def\rb{\rbrack}

\def\llb{\left\lbrack}
\def\rrb{\right\rbrack}

\def\lcurl{\left\{}
\def\rcurl{\right\}}
\def\({\left(}
\def\){\right)}
\def\lskip{\vskip\baselineskip\vskip-\parskip\noindent}
\def\mskp{\par\vskip 0.3cm \par\noindent}

\def\bc{\begin{center}}
\def\ec{\end{center}}

\newcommand\partder[2]{{{\partial {#1}}\over{\partial {#2}}}}

\newcommand\sbr[2]{\left\lbrack\,{#1}\, ,\,{#2}\,\right\rbrack} 
\newcommand\Sbr[2]{\Bigl\lbrack\,{#1}\, ,\,{#2}\,\Bigr\rbrack} 
 
\def\a{\alpha}
\def\b{\beta}

\def\d{\delta}

\def\vareps{\varepsilon}
\def\g{\gamma}

\def\l{\lambda}

\def\o{\over}

\def\vp{\varphi}
\def\P{\Phi}
\def\pa{\partial}

\def\bpa{{\bar \partial}}
\def\pr{\prime}

\def\t{\tau}

\def\wti{\widetilde}

\newcommand\BDet[5]{\det_{{#1}}\left\Vert\begin{array}{cc}  
{#2} & {#3} \\ {#4} & {#5} \end{array} \right\Vert}   

\def\cL{{\cal L}}
\def\cM{{\cal M}}

\def\cW{{\cal W}}

\def\mark{\noindent{\bf Remark.}\quad}

\def\proof{\par{\it Proof}. \ignorespaces} \def\endproof{{$\Box$}\par}

\newcommand\DB{{Darboux-B\"{a}cklund}~}
\def\bt{{\bar t}}
\def\pai{\partial^{-1}\!\!}
\def\bD{{\bar D}}
\def\bpai{{\bar \partial}^{-1}\!\!}
\def\bcL{{\bar {\cal L}}}
\def\bP{{\bar \Phi}}
\def\bPsi{{\bar \Psi}}

\newcommand\st[2]{\stackrel{(#1 )}{#2}}
\newcommand\stt[1]{\stackrel{(#1 )}{t}}
\newcommand\stta[2]{\stackrel{(#1 )}{t_{#2}}}
\newcommand\spa[1]{\stackrel{(#1 )}{\partial}}
\newcommand\spaa[2]{\stackrel{(#1 )}{\partial_{#2}}}
\newcommand\spai[1]{\stackrel{(#1 )}{\partial^{-1}}\!\!}
\newcommand\sP[2]{\stackrel{(#1 )}{\Phi_{#2}}}
\newcommand\sPsi[2]{\stackrel{(#1 )}{\Psi_{#2}}}

\newcommand{\ct}[1]{\cite{#1}}
\newcommand{\bi}[1]{\bibitem{#1}}
\newcommand\PRL[3]{{\sl Phys. Rev. Lett.} {\bf#1} (#2) #3}

\newcommand\CMP[3]{{\sl Commun. Math. Phys.} {\bf #1} (#2) #3}

\newcommand\PLA[3]{{\sl Phys. Lett.} {\bf #1A} (#2) #3}

\newcommand\PHSA[3]{{\sl Physica} {\bf A#1} (#2) #3}

\begin{document}
\vspace*{-1.5cm}
\noindent
{\sl solv-int/9904024} \hfill{BGU-99/20/Apr-PH}\\

\begin{center}
{\large {\bf Multi-Component Matrix KP Hierarchies as Symmetry-Enhanced 
Scalar KP Hierarchies and Their Darboux-B\"acklund Solutions}}  
\end{center}    
 
\begin{center}
H. Aratyn${}^1$\footnotetext[1]{Department of Physics, University of 
Illinois at Chicago, 845 W. Taylor St., Chicago, IL 60607-7059, U.S.A.;
e-mail: aratyn@uic.edu}, 
E. Nissimov${}^{2,3}$ and S. Pacheva${}^{2,3}$
\footnotetext[2]{Institute of Nuclear Research and Nuclear Energy,
Boul. Tsarigradsko Chausee 72, BG-1784 $\;$Sofia, Bulgaria;
e-mail: nissimov@inrne.bas.bg , svetlana@inrne.bas.bg}
\footnotetext[3]{Department of Physics, Ben-Gurion University of the Negev,
Box 653, IL-84105 $\;$Beer Sheva, Israel; e-mail: emil@bgumail.bgu.ac.il, 
svetlana@bgumail.bgu.ac.il}
\end{center} 

\begin{abstract}
We show that any multi-component matrix KP hierarchy is equivalent to
the standard one-component (scalar) KP hierarchy endowed with a special 
infinite set of abelian additional symmetries, generated by squared 
eigenfunction potentials. This allows to employ a special version of the
familiar Darboux-B\"acklund transformation techniques within the ordinary
scalar KP hierarchy in the Sato formulation for a systematic derivation of 
explicit {\em multiple-}Wronskian tau-function solutions of all 
multi-component matrix KP hierarchies.
\end{abstract}  

\lskip
{\bf 1. Introduction. Background on the KP Hierarchy and Ghosts Symmetries.}

Multi-component generalizations of Kadomtsev-Petviashvili (KP) hierarchy of
integrable nonlinear soliton equations \ct{multi-KP} attract a lot of
interest both from physical and mathematical point of view. They are known to
contain such physically relevant nonlinear integrable systems as
Davey-Stewartson, two-dimensional Toda lattice and three-wave resonant
interaction ones \ct{DS}. On the other hand, multi-component KP hierarchies 
turn out to be intimately connected to classical geometry of conjugate nets
and the classification problem of Hamiltonian systems of hydrodynamical type
\ct{multi-KP-geom}.

There exist several equivalent formulations of multi-component KP hierarchies:
matrix pseudo-differential operator (Sato) formulation; tau-function approach
via matrix Hirota bilinear identities; multi-component free 
fermion formulation. Here we will offer yet another approach. 
Namely, we will show that any multi-component $N\!\times\! N$ matrix KP 
hierarchy \ct{multi-KP} is equivalent to the standard one-component (scalar) 
KP hierarchy endowed with $N\!-\! 1$ copies of mutually commuting 
infinite-dimensional algebras of abelian additional (``ghost'') symmetries, 
generated by squared eigenfunction potentials of the initial hierarchy 
(see definition in \rf{SEP-def} below). 
The latter ``ghost'' symmetry enhanced scalar KP hierarchy will be called 
``multiple-KP hierarchy''. The principal advantage of multiple-KP hierarchy
formulation over the standard Sato matrix pseudo-differential operator
formulation lies in the fact that the former allows to use a special version 
of the well-known Darboux-B\"acklund (DB) transformation techniques within
the ordinary scalar KP hierarchy in the Sato formulation for a systematic 
derivation of soliton-like {\em multiple-}Wronskian tau-function 
solutions of the multi-component KP hierarchies (Section 5 below).

The starting point of our presentation is the pseudo-differential Lax 
operator $\cL$ of scalar KP hierarchy obeying KP evolution equations w.r.t. 
the multi-time $(t) \equiv (t_1 \equiv x, t_2 ,\ldots )$ (for notations and 
review, see \ct{ldickey}) :
\be
\cL = D + \sum_{i=1}^{\infty} u_i D^{-i} \qquad ; \qquad
\partder{\cL}{t_l} = \Sbr{\(\cL^{l}\)_{+}}{\cL} \quad , \; \;
l = 1, 2, \ldots
\lab{lax-eq}
\ee
The symbol $D$ stands for the differential operator $\pa/\pa x$, whereas
$\pa \equiv \pa_x$ will denote derivative of a function. The subscripts
$(\pm)$ indicate purely differential/pseudo-differential part of the
corresponding operator.
Equivalently, one can represent Eq.\rf{lax-eq} in terms of the 
dressing operator $W$ whose pseudo-differential series are 
expressed in terms of the so called tau-function $\t (t)$ :
\be
\cL = W D W^{-1} \quad ,\quad \partder{W}{t_l} = - \(\cL^l\)_{-} W \quad ,
\quad W = \sum_{n=0}^{\infty} \frac{p_n \( - [\pa]\)\t (t)}{\t (t)} D^{-n}
\lab{W-main}
\ee
with the notation: $[y] \equiv \( y_1,  y_2/2 , y_3/3 ,\ldots \)$ for any
multi-variable $(y) \equiv \( y_1 ,y_2 ,y_3 ,\ldots \)$, in particular
$(\pa ) \equiv \(\pa/\pa t_1 ,\pa/\pa t_2 ,\ldots \)$, and with $p_k (y)$
being the Schur polynomials.
The tau-function is related to the Lax operator as (below ``Res'' denotes
the coefficient in front of $D^{-1}$) :
\be
\pa_x \partder{}{t_l}\ln \t (t) = {\rm Res} \cL^{l} \quad ,\quad
\t (t) \longrightarrow \t (t) \, e^{\sum_{l=1}^\infty c_l\, t_l}
\lab{tau-L}
\ee
The second relation \rf{tau-L} tells that $\t (t)$ is defined up to an
exponential linear w.r.t. $(t)$.

In the present approach a basic notion is that of (adjoint) eigenfunctions
(EF's) $\P (t),\, \Psi (t)$ of the scalar KP hierarchy satisfying :
\be
\partder{\Phi}{t_k} = \cL^{k}_{+}\bigl( \Phi\bigr) \qquad; \qquad
\partder{\Psi}{t_k} = - \(\cL^{*} \)^{k}_{+}\bigl( \Psi\bigr)
\lab{eigenlax}
\ee

Throughout this paper we will rely on an important tool provided by the
spectral representation of EF's \ct{ridge}.
The spectral representation is equivalent to the following statement:
$\P$ and $\Psi$ are (adjoint) EF's if and only if they obey the
integral representation:
\br
\P (t) &=&  \int\!\! dz\, {e^{\xi (t-t^{\pr} ,z )}\o z} 
{ \t (t - [z^{-1}]) \t (t^{\pr} + [z^{-1}]) \o \t (t)  \t (t^{\pr} )  } \, 
  \P \( t^{\pr} + [z^{-1}]\) 
\lab{spec1t-a}\\
\Psi (t) &=& \int\!\! d z\, {e^{\xi (t^{\pr}-t ,z )}\o z} 
{ \t (t + [z^{-1}]) \t (t^{\pr} - [z^{-1}]) \o \t (t)  \t (t^{\pr} )  } \,
\Psi (t^{\pr} - [z^{-1}])
\lab{spec2t-a}
\er
where $\int\!\! dz$ denotes normalized contour integral around origin.

Further crucial notion to be employed in the present contsruction is the
so called squared eigenfunction potential (SEP) $S\(\P ,\Psi\)$ of arbitrary
pair of an EF $\P$ and an adjoint EF $\Psi$ 
\ct{oevela} (cf. also \ct{ridge}) :
\be
S\(\P ,\Psi\) = \pai\(\P \,\Psi\) \quad ,\quad 
{\pa \o \pa t_n} S \( \P (t), \Psi (t) \) = 
{\rm Res} \( D^{-1} \Psi (\cL^n)_{+} \P D^{-1} \) \;\; ,\; n=1,2,\ldots
\lab{SEP-def}
\ee
The flow equations above fix the ambiguity in applying the inverse
derivative $\pai$ in the SEP definition up to an overall trivial constant.
In what follows $\pai$ will appear always in the form of SEP's 
(first Eq.\rf{SEP-def}).

Consider now an infinite system of independent (adjoint) EF's
$\lcurl \P_j ,\Psi_j\rcurl_{j=1}^{\infty}$ of the standard one-component KP 
hierarchy Lax operator $\cL$ and define the following infinite set of the 
additional ``ghost'' symmetry flows \ct{hungry}\foot{Similar additional
symmetries have been also considered in the particular case of constrained
KP hierarchies \ct{EOR95}.}:
\be
\partder{}{\bt_s} \cL = \Sbr{\cM_s}{\cL} \qquad ,\qquad
\cM_s = \sum_{j=1}^s \P_{s-j+1} D^{-1} \Psi_j 
\lab{ghost-s}
\ee
\be
\partder{}{\bt_s} \P_k = \sum_{j=1}^s \P_{s-j+1} \pai\(\Psi_j\P_k\) -
\P_{k+s} \quad ; \quad
\partder{}{\bt_s} \Psi_k = \sum_{j=1}^s \Psi_j \pai\(\P_{s-j+1}\Psi_k\)
+ \Psi_{k+s} 
\lab{M-s-eigenf}
\ee
\be
\partder{}{\bt_s} F = \sum_{j=1}^s \P_{s-j+1} \pai\( F \Psi_j \) \quad ;\quad
\partder{}{\bt_s} F^\ast = \sum_{j=1}^s \Psi_j \pai\(\P_{s-j+1} F^\ast\)
\lab{M-s-generic}
\ee
where $s,k =1,2,\ldots$ , and $F,\, F^{*}$ denote generic (adjoint)
EF's which do not belong to the ``ghost''
symmetry generating set $\lcurl \P_j ,\Psi_j\rcurl_{j=1}^{\infty}$.

It is easy to show that the ``ghost'' symmetry flows
$\pa / \pa {\bt_s}$ from Eqs.\rf{ghost-s}-\rf{M-s-generic} commute, 
namely, that the $\pa$-pseudo-differential
operators $\cM_s $ \rf{ghost-s} satisfy the zero-curvature equations:
$ \pa \cM_r / \pa {\bt_s} - \pa \cM_s /\pa \bt_r - \sbr{\cM_s}{\cM_r} = 0$.

Let us consider the (adjoint)-EF Eqs.\rf{eigenlax} for $k\! =\! 2$ 
and $\P\! =\! \P_1,\, \Psi\! =\! \Psi_1$ (the pair generating the lowest 
$s\! =\! 1$ ``ghost'' symmetry flows \rf{ghost-s}--\rf{M-s-generic}) together 
with their $s\! =\! 2$ ``ghost'' symmetry flow Eqs.\rf{M-s-eigenf} ({\sl i.e.},
$s\! =\! 2,\, k\! =\! 1$). The latter system can be written in the form:
\br
\partder{}{t_2} \P_1 = \(\pa^2 + 2u_1\) \P_1 \quad ,\quad
\partder{}{\bt_2} \P_1 = \llb - \bpa^2 + 2\bpa\(\pai\(\P_1 \Psi_1\)\)\rrb \P_1
\lab{pa-bpa-2-P1} \\
\partder{}{t_2} \Psi_1 = - \(\pa^2 + 2u_1\) \Psi_1 \quad ,\quad
\partder{}{\bt_2} \Psi_1 = \llb \bpa^2 - 2\bpa\(\pai\(\P_1 \Psi_1\)\)\rrb \Psi_1
\lab{pa-bpa-2-Psi1}
\er
where $ \bpa \equiv \pa/ \pa {\bt_1} $. Introducing new notations:
\be
Q \equiv u_1 - 2\(\P_1 \Psi_1\) - \bpa\(\pai\(\P_1 \Psi_1\)\) \quad ;
\quad -iT = t_2 - \bt_2 \;\; ,\;\; X = t_1 + \bt_1 \;\; ,\;\; 
Y = t_1 - \bt_1
\lab{new-vars}
\ee
the system \rf{pa-bpa-2-P1}--\rf{pa-bpa-2-Psi1} acquires the following form:
\be
i\pa_T \P_1 = \Bigl(\pa_X^2 + \pa_Y^2\Bigr) \P_1 +
2\(\P_1 \Psi_1\)\,\P_1 + Q\,\P_1  \quad ,\quad
\Bigl(\pa_X^2 - \pa_Y^2\Bigr) Q + 4\pa_X^2 \(\P_1 \Psi_1\) = 0  
\lab{DS-eqs} 
\ee
\be
- i\pa_T \Psi_1 = \Bigl(\pa_X^2 + \pa_Y^2\Bigr) \Psi_1 +
2\(\P_1 \Psi_1\)\,\Psi_1 + Q\,\Psi_1      \nonu
\ee
which is nothing but the standard Davey-Stewartson (DS) system 
\ct{DS}\foot{The fact that DS system is contained within the ``ghost''
symmetry enhanced scalar KP hierarchy 
\rf{lax-eq}--\rf{W-main},\rf{ghost-s}--\rf{M-s-eigenf} was first pointed out 
in \ct{Hisakado}.}.
Thus, we succeeded to express solutions of DS system in terms of a pair of
(adjoint) EF's of ordinary one-component KP hierarchy supplemented with two
additional ``ghost'' symmetry flows ($\partder{}{\bt_1}$ and 
$\partder{}{\bt_2}$ in the notations of \rf{ghost-s}--\rf{M-s-eigenf}).

Consider now the $\t$-function of $\cL$ \rf{W-main}
and let us act with $\partder{}{\bt_s}$ on both sides of \rf{tau-L} obtaining~
$\partder{}{\bt_s} \ln \t = - \sum_{j=1}^s \pai \(\P_{s-j+1} \Psi_j \)$
where we used \rf{ghost-s} as well as the $t_r$-flow eqs.
$\partder{}{t_r}\cM_s = {\Sbr{\cL^r_{+}}{\cM_s}}_{-}$.
With the help of the well-known recurrence relation for the Schur polynomials:
$s p_s\(-[\bpa]\)=
\sum_{k=1}^s \Bigl(-\partder{}{\bt_k}\Bigr) p_{s-k}\(-[\bpa]\)$,
we find the following important identities \ct{hungry,zim-dkp}
relating the tau-function with the ``ghost'' symmetry generating
(adjoint) EF's (here $s,k,j = 1,2,\ldots $):
\be
\frac{p_s \( -[\bpa]\) \(\P_1 \t \)}{\t} = \P_{s+1}   \;\; ,\;\;
\frac{p_s \( [\bpa]\) \(\Psi_1 \t\)}{\t} = \Psi_{s+1}  \;\; ; \;\; 
\pai\(\P_k\Psi_j\) = 
\sum_{l=0}^{j-1} \frac{p_{k+l}(-[\bpa])\, p_{j-l+1}([\bpa])\,\t}{\t}
\lab{p-s-tau-P-Psi}
\ee
\noindent
{\bf 2. Double-KP System and Its Equivalence to Two-Component KP Hierarchy}

In refs.\ct{hungry,zim-dkp} we have shown that the ``ghost'' symmetry flows
from Eqs.\rf{ghost-s}-\rf{M-s-generic} admit their own Lax representation
in terms of a $\bD \equiv \pa/\pa {\bar x} \equiv \pa/\pa \bt_1$ 
pseudo-differential Lax operator $\bcL$ w.r.t. multi-time 
$({\bar t}) \equiv ({\bar t}_1 \equiv {\bar x}, {\bar t}_2 ,\ldots )$ :
\br
\bcL \equiv {\overline {\cW}}\, {\bD}\, {\overline {\cW}}^{-1} =
{\bD} + \sum_{i=1}^\infty {\bar u}_i {\bD}^{-i}  \quad ,\quad
\partder{\bcL}{\bt_s} = \Sbr{{\bcL}^s_{+}}{{\bcL}}
\lab{bar-L} \\
{\overline {\cW}} =
1 + \sum_{j=1}^\infty \frac{p_j (-[\bpa])\t (t,\bt )}{\t (t,\bt )}{\bD}^{-j} 
\quad ,\quad  \frac{p_j (-[\bpa])\t}{\t} = \pai\(\P_j \Psi_1\)
\lab{bar-W}
\er
where $\P_j,\, \Psi_1$ are the ``ghost'' symmetry generating (adjoint)
EF's of the original KP hierarchy \rf{lax-eq}, which we denote
as $K\! P_1$. Accordingly, we will denote the ``ghost''KP system 
\rf{bar-L}--\rf{bar-W} as $K\! P_2$. Let us stress that
the tau-function $\t = \t (t,\bt )$ of the $K\! P_1$ hierarchy 
\rf{lax-eq}--\rf{W-main}, with time evolution parameters $(t)$ and 
$(\bt )= fixed$, is simultaneously tau-function of $K\! P_2$ 
hierarchy \rf{bar-L}--\rf{bar-W} with time evolution parameters $(\bt )$ and 
$(t) = fixed$. 

Next, we have found an infinite system of (adjoint) EF's
$\lcurl \bP_k,\bPsi_k \rcurl_{k=1}^\infty$ of $K\! P_2$ system
({\sl i.e.}, obeying Eqs.\rf{eigenlax} with all quantities
replaced with the ``barred'' ones) :
\be
\bP_k = \frac{p_{k-1}(-[\pa])\(\Psi_1 \t\)}{\t}  \quad ,\quad
\bPsi_k = \frac{p_{k-1}([\pa])\(\P_1 \t\)}{\t}
\lab{dual-eigenf}
\ee
which generate ``ghost'' symmetry flows for $K\! P_2$ \rf{bar-L}
analogous to \rf{ghost-s}--\rf{M-s-generic} such that the corresponding
``ghost'' symmetry flow parameters coincide with the isospectral flow
parameters $(t)$ of the original $K\! P_1$ system \rf{lax-eq}. In particular,
we note from Eqs.\rf{dual-eigenf} that $\bP_1 = \Psi_1$ and $\bPsi_1 = \P_1$.
Also, for any generic (adjoint) EF's $F$, $F^\ast$ of $K\! P_1$ the SEP
functions: 
\be
{\bar F} \equiv \pai\( F\,\Psi_1\) \quad ,\quad
{\bar F}^\ast \equiv \pai\(\P_1 F^\ast\)
\lab{SEP-EF}
\ee
are, respectively, an EF and adjoint EF of $K\! P_2$ \ct{hungry,zim-dkp}.

Both $K\! P_1$ (original KP hierarchy) together with $K\! P_2$ (``ghost''
symmetry flows' KP hierarchy) form a new larger hierarchy called 
{\em double-KP} \ct{hungry,zim-dkp} possessing the property of ``duality'' 
symmetry, {\sl i.e.}, symmetry under interchanging the r\^oles of $K\! P_1$
and $K\! P_2$. It is defined as follows:
\be
(t) \equiv (\stt{1}) \;\; ,\;\; (\bt ) \equiv (\stt{2}) \;\; ,\;\;
(\stt{\a}) = \bigl(\stta{\a}{1},\stta{\a}{2},\ldots \bigr)
\quad ,\quad  \cL \equiv \st{1}{\cL} \quad ,\quad  \bcL \equiv \st{2}{\cL}  
\lab{symm-notation} 
\ee
\be
\st{\a}{\cL} = \st{\a}{W} \st{\a}{D} \st{\a}{W^{-1}} \quad ,\quad
\st{\a}{W} = \sum_{j=0}^\infty \frac{p_j (-[\spa{\a}])\t}{\t} \st{\a}{D^{-j}}
\quad ,\quad   \st{\a}{D} \equiv \pa/\pa \stta{\a}{1} 
\lab{Lax-W-a}
\ee
\be
\P_j (t,\bt ) \equiv \sP{12}{j} (\stt{1}\!,\!\stt{2}) \;\; ,\;\;
\Psi_j (t,\bt ) \equiv \sPsi{12}{j} (\stt{1}\!,\!\stt{2})  \;\; ,\;\;
\bP_j (t,\bt ) \equiv \sP{21}{j} (\stt{1}\!,\!\stt{2})  \;\; ,\;\;
\bPsi_j (t,\bt ) \equiv \sPsi{21}{j} (\stt{1}\!,\!\stt{2}) 
\lab{sP-sPsi-def}
\ee
\be
\sP{\a\b}{j} = \vareps_{\a\b} \frac{p_{j-1}(-[\spa{\b}])\t_{\a\b}}{\t}
\quad ,\quad
\sPsi{\a\b}{j}= \vareps_{\b\a} \frac{p_{j-1}([\spa{\b}])\t_{\b\a}}{\t}
\quad ,\quad \vareps_{\a\b} = \pm 1 \;\; {\rm for}\; \a \leq \b \; ,\; \a > \b
\lab{sP-sPsi-tau}
\ee
where in \rf{sP-sPsi-tau} we have introduced new tau-functions:
\be
\t_{\a\b} = \vareps_{\a\b}\,\t\!\!\sP{\a\b}{1} = 
\vareps_{\b\a}\,\t\!\!\sPsi{\b\a}{1}
\lab{tau-ab-def}
\ee
Here the indices $(\a,\b)$ are taking values $\a,\b = 1,2$ , $\a \neq \b$ ,
and Eqs.\rf{p-s-tau-P-Psi},\rf{dual-eigenf} have been taken into 
account.

Accordingly, 
Eqs.\rf{lax-eq},\rf{bar-L},\rf{eigenlax},\rf{ghost-s}--\rf{M-s-generic} and 
their ``duals'' (for $\bP_k,\,\bPsi_k$) 
can be rewritten in a manifestly ``duality''-symmetric form:
\be
\spaa{\a}{s} \st{\a}{\cL} = 
\Sbr{\Bigl(\st{\a}{\cL}\Bigr)^s_{+}}{\st{\a}{\cL}} \quad ,\quad
\spaa{\b}{s} \st{\a}{\cL} = 
\Sbr{\st{\a\b}{\cM_s}}{\st{\a}{\cL}}  \quad ,\quad
\spaa{\a}{s} \equiv \partder{}{\stta{\a}{s}} \quad ,\quad 
\st{\a\b}{\cM_s} \equiv 
\sum_{j=1}^s \st{\a\b}{\P}\!\!\!\!{}_{s-j+1} \st{\a}{D^{-1}} \sPsi{\a\b}{j}
\lab{Lax-eqs-ab} 
\ee
\be
\spaa{\a}{s} \sP{\a\b}{k} = \Bigl(\st{\a}{\cL}\Bigr)^s_{+} 
\bigl(\sP{\a\b}{k}\bigr) \;\; ,\;\;
\spaa{\a}{s} \st{\a}{F} = \Bigl(\st{\a}{\cL}\Bigr)^s_{+} 
\bigl(\st{\a}{F}\bigr) \quad ;\quad
\spaa{\a}{s} \sPsi{\a\b}{k} = - \Bigl(\st{\a}{\cL^\ast}\Bigr)^s_{+} 
\bigl(\sPsi{\a\b}{k}\bigr) \;\;,\;\;
\spaa{\a}{s} \st{\a}{F^\ast} = - \Bigl(\st{\a}{\cL^\ast}\Bigr)^s_{+} 
\bigl(\st{\a}{F^\ast}\bigr)
\lab{eigenlax-a}
\ee
\be
\spaa{\b}{s} \sP{\a\b}{k} = 
\sum_{j=1}^s \st{\a\b}{\P}\!\!\!\!{}_{s-j+1} \spai{\a}\Bigl(\sPsi{\a\b}{j}\sP{\a\b}{k}\Bigr)
- \st{\a\b}{\P}\!\!\!\!{}_{k+s} \quad ,\quad
\spaa{\b}{s} \sPsi{\a\b}{k} = 
\sum_{j=1}^s \sPsi{\a\b}{j} \spai{\a}\Bigl(\st{\a\b}{\P}\!\!\!\!{}_{s-j+1}\sPsi{\a\b}{k}\Bigr)
+ \st{\a\b}{\Psi}\!\!\!\!{}_{k+s} 
\lab{M-s-ab}
\ee
\be
\spaa{\b}{s} \st{\a}{F} = 
\sum_{j=1}^s \st{\a\b}{\P}\!\!\!\!{}_{s-j+1} \spai{\a}\Bigl(\sPsi{\a\b}{j} \st{\a}{F}\Bigr)
\quad ,\quad
\spaa{\b}{s} \st{\a}{F^\ast} = 
\sum_{j=1}^s \sPsi{\a\b}{j}\spai{\a}\Bigl(\st{\a\b}{\P}\!\!\!\!{}_{s-j+1}\st{\a}{F^\ast}\Bigr)
\lab{M-generic-ab}
\ee
where again $\a,\b\! =\! 1,2$ , $\a\! \neq\! \b$ ,
$\spai{\a}\;\equiv \bigl(\st{\a}{\pa_1}\bigr)^{-1}$ and $s\! =\! 1,2,\ldots$ .
In \rf{M-generic-ab} and \rf{eigenlax-a} $\st{\a}{F},\,\st{\a}{F^\ast}$ denote 
generic (adjoint) EF's of $\st{\a}{\cL}$, {\sl i.e.}, such that they do not
belong to the sets $\bigl\{\sP{\a\b}{k},\,\sPsi{\a\b}{k}\bigr\}$. 

In \ct{zim-dkp} we have shown the equivalence of double-KP hierarchy with
the two-component (matrix) Sato KP hierarchy \ct{multi-KP}, originally
formulated within the matrix pseudo-differential Lax formalism, which can be
equivalently described by three tau-functions $\t_{11}, \t_{12}, \t_{21}$
depending on two sets of multi-time variables $(t),\, ({\bar t})$ and obeying 
the $2\!\times\! 2$ matrix Hirota bilinear identities (see Eq.\rf{HBI-ab} 
below). The proof proceeds by identifying two-component KP tau-functions with 
the tau-functions of double-KP hierarchy 
$\t_{11}= \t,\,\t_{12}= \t\P_1,\,\t_{21}= - \t\Psi_1$ as
in \rf{Lax-W-a}--\rf{tau-ab-def}.
\mskp
{\bf 3. Generalization to Multi-Component KP Hierarchies}

Let us turn our attention to arbitrary $N$-component matrix KP hierarchies 
\ct{multi-KP} ($N \geq 2$). They can be equivalently characterized by the set
of Hirota bilinear identities: 
\br
\sum_{\g =1}^N \vareps_{\a\g}\, \vareps_{\b\g}\int\! dz \,
z^{\d_{\a\g} + \d_{\b\g} -2}\, e^{\xi (\st{\g}{t}-\st{\g}{t^\pr},z)}
\t_{\a\g}\bigl(\ldots ,\st{\g}{t}-[z^{-1}],\ldots \bigr)\,
\t_{\g\b}\bigl(\ldots ,\st{\g}{t^\pr}+[z^{-1}],\ldots \bigr) = 0
\lab{HBI-ab}
\er
for $N(N-1)+1$ tau-functions ~$\t_{\a\a}\!\equiv \!\t$ and
$\t_{\a\b} \, (\a\!\neq\!\b )$,
where now the indices $\a,\b,\g = 1,\ldots ,N$, $\d_{\a\b}$ are Kronecker
symbols and $\vareps_{\a\b}$ are the same as in \rf{sP-sPsi-def}. Also, we
are using the standard notation $\xi (t,z) \equiv \sum_{l=1}^\infty t_l z^l$.
Let us recall that \rf{HBI-ab} contain the following interesting systems of
non-linear equations:
\be
\spaa{\a}{1}\spaa{\b}{1} \ln\t = \frac{\t_{\a\b}}{\t}\, \frac{\t_{\b\a}}{\t}
\quad ,\quad
\spaa{\g}{1}\(\frac{\t_{\a\b}}{\t}\) =
\vareps_{\a\b}\,\vareps_{\a\g}\,\vareps_{\g\b}\,
\frac{\t_{\a\g}}{\t}\, \frac{\t_{\g\b}}{\t}
\quad ,\quad   \a \neq \b \neq \g
\lab{N-wave-tau}
\ee
the second one being the so called $N^\pr$-wave system ($N^\pr = N(N-1)/2$). 

Our main statement is that $N$-component KP hierarchy defined by \rf{HBI-ab}
is equivalent to the {\em multiple-KP} hierarchy defined by
Eqs.\rf{Lax-W-a},\rf{sP-sPsi-tau}--\rf{M-generic-ab} where now the
indices $\a,\b$ take values $\a,\b = 1,\ldots ,N$. In other words, this
multiple-KP hierarchy consists of $N$ ordinary one-component KP hierarchies
$K\! P_{\a}$ ($\a\!=1,\ldots ,N$) given by Lax operators $\st{\a}{\cL}$ 
\rf{Lax-W-a} in different spaces and generating isospectral flows 
$\spaa{\a}{s}$ w.r.t. different sets of evolution parameters $(\st{\a}{t})$, 
such that the flows $\spaa{\a}{s}$ act on the rest of KP subsystems 
$K\! P_{\b}$ ($\b\!\neq\!\a$) as ``ghost'' symmetry flows. In particular, 
Eqs.\rf{M-generic-ab} apply also for $\st{\a}{F}\! =\!\sP{\a\g}{k}$ and 
$\st{\a}{F^\ast}\! =\!\sPsi{\a\g}{k}$ with $\g\!\neq\!\b\!\neq\!\a$ :
\be
\spaa{\b}{s} \sP{\a\g}{k} = 
\sum_{j=1}^s \st{\a\b}{\P}\!\!\!\!{}_{s-j+1} \spai{\a}\Bigl(\sPsi{\a\b}{j} 
\sP{\a\g}{k}\Bigr)      \quad ,\quad
\spaa{\b}{s} \sPsi{\a\g}{k} = \sum_{j=1}^s \sPsi{\a\b}{j}
\spai{\a}\Bigl(\st{\a\b}{\P}\!\!\!\!{}_{s-j+1}\sPsi{\a\g}{k}\Bigr)
\lab{M-s-ag}
\ee
since $\sP{\a\g}{k},\sPsi{\a\g}{k}$ are generic (adjoint) EF's
of $\st{\a}{\cL}$ w.r.t. ``ghost'' symmetry flows $\spaa{\b}{s}$ when
$\g\!\neq\!\b\!\neq\!\a$.

The detailed proof follows closely the pattern of the proof for the 
$N=2$ component KP case \ct{zim-dkp}, making heavy use of the spectral
representation identities \rf{spec1t-a}--\rf{spec2t-a}, and will be given in a
separate paper.

Here we will only present an additional important property of multiple-KP
system \rf{Lax-W-a},\rf{sP-sPsi-tau}--\rf{M-generic-ab},\rf{M-s-ag}, which
appears only in the $N\geq 3$ cases. Namely, for any $\a\!\neq\!\b\!\neq\!\g$
the following identities hold among (adjoint) EF's:
\be
\spaa{\g}{1} \sP{\a\b}{k} = \sP{\a\g}{1} \sP{\g\b}{k} \quad ,\quad
\spaa{\g}{1} \sPsi{\a\b}{k} = \sPsi{\a\g}{1} \sPsi{\g\b}{k} 
\lab{N-wave-EF}
\ee
In particular, taking $k\!=\! 1$ in the first Eq.\rf{N-wave-EF} and
taking into account \rf{tau-ab-def}, we see that the latter coincides with the
$N^\pr$-wave system \rf{N-wave-tau}. In other words, the $N^\pr$-wave 
system is reformulated entirely in terms of EF's of an underlying ``ghost''
symmetry enhanced ordinary KP hierarchy. Moreover, it is easy to check that
\rf{N-wave-EF} are compatible with the ``ghost'' flow Eqs.\rf{M-s-ag},
and in fact \rf{N-wave-EF} are equivalent to \rf{M-s-ag}.
\mskp
{\bf 4. \DB Orbits of Multi-Component KP Hierarchies}

We will consider DB transformations in their form appropriate for the Sato
formulation of KP hierarchies \ct{oevela}. Namely, DB transformations are
pseudo-differential operator ``gauge'' transformations of the pertinent Lax
operators given in terms of (adjoint) EF's \rf{eigenlax}.

Let us temporarily return to the simpler case of double-KP hierarchy.
In our construction a very instrumental r\^{o}le will be played by the
following {\em non-standard} orbit of successive DB transformations for the
original $K\! P_1$ system \rf{lax-eq} 
($\cL \equiv \cL (n)\, ,\, \sP{12}{j}\equiv\P_j \equiv \P_j^{(n)},$ etc.) :
\be
\cL (n+1) = T(n) \cL (n) T^{-1} (n) \quad ,\quad
T(n) = \P_1 D \P_1^{-1} \equiv \P_1^{(n)} D {\P_1^{(n)}}^{\, -1} 
\lab{DB-L}
\ee
\be
\P_l^{(n+1)} = \P_1^{(n)} \pa\!\(\frac{\P_{l+1}^{(n)}}{\P_1^{(n)}}\) 
\; ,\; l \geq 1 \; ;
\;\;\; \Psi_1^{(n+1)} = \frac{1}{\P_1^{(n)}} \;\; ,\;\;
\Psi_j^{(n+1)} = - \frac{1}{\P_1^{(n)}} \pai \(\P_1^{(n)} \Psi_{j-1}^{(n)}\) 
\; ,\; j \geq 2
\lab{DB-eigenf}
\ee
\be
F^{(n+1)} = \P_1^{(n)} \pa\(\frac{F^{(n)}}{\P_1^{(n)}}\) \quad ,\quad
{F^\ast}^{(n+1)} = - \frac{1}{\P_1^{(n)}} \pai \(\P_1^{(n)} {F^\ast}^{(n)}\)
\lab{DB-F}
\ee    
for transformations in ``positive'' direction, as well as adjoint DB
transformations, {\sl i.e.}, transformations in ``negative'' direction: 
\be
\cL (n-1) = {\widehat T}^{\ast\, -1} (n) \cL (n) {\widehat T}^{\ast} (n) 
\quad ,\quad    {\widehat T} (n) = \Psi_1 D \Psi_1^{-1} \equiv 
\Psi_1^{(n)} D {\Psi_1^{(n)}}^{\, -1} 
\lab{adjDB-L}
\ee
\be
\P_1^{(n-1)} = \frac{1}{\Psi_1^{(n)}} \;\; , \;\;
\P_l^{(n-1)} = \frac{1}{\Psi_1^{(n)}} \pai \(\Psi_1^{(n)} \P_{l-1}^{(n)}\) 
\;\; ,\; l \geq 2   \; ;\;\;
\Psi_j^{(n-1)} = - \Psi_1^{(n)} \pa\!\(\frac{\Psi_{j+1}^{(n)}}{\Psi_1^{(n)}}\)
\;\; , \; j \geq 1
\lab{adjDB-eigenf}
\ee
\be
F^{(n-1)} = \frac{1}{\Psi_1^{(n)}} \pai \(\Psi_1^{(n)} F^{(n)}\) \quad,\quad
{F^\ast}^{(n-1)} = - \Psi_1^{(n)} \pa \(\frac{{F^\ast}^{(n)}}{\Psi_1^{(n)}}\)
\lab{adjDB-F}
\ee
In what follows, the DB ``site'' index $(n)$  will be skipped
for brevity whenever this would not lead to ambiguities. Note that under the
above DB transformations, the tau-function $\t \equiv \t^{(n)}$ transforms as:
$\t^{(n+1)} = \P_1 \t \equiv \P_1^{(n)} \t^{(n)}$ ,
$\t^{(n-1)} = -\Psi_1 \t \equiv -\Psi_1^{(n)} \t^{(n)}$.
\mskp
\mark
Let us stress the non-canonical form of the (adjoint) DB transformations
\rf{DB-eigenf},\rf{adjDB-eigenf} on the ``ghost'' symmetry generating 
(adjoint) EF's.  
On the other hand, for generic (adjoint) EF's $F, F^\ast$ the 
(adjoint) DB transformations \rf{DB-F},\rf{adjDB-F} read as 
usual \ct{oevela}.
\mskp

The  crucial property of the above DB orbit of the original one-component
$K\! P_1$ hierarchy is that it induces an orbit of DB transformations for the
whole double-KP system \rf{Lax-W-a}--\rf{M-generic-ab}. More precisely,
as shown in \ct{hungry,zim-dkp}, ``ghost'' symmetries \rf{ghost-s} commute
with DB transformations \rf{DB-L}--\rf{adjDB-F} of the $K\! P_1$ hierarchy,
and induce the following DB transformations on the ``ghost''
$K\! P_2$ hierarchy (recall from Eqs.\rf{dual-eigenf}, that
$\bP_1^{(n)} = \Psi_1^{(n)}\, ,\, \bPsi_1^{(n)} = \P_1^{(n)}$) :
\be
\bcL(n+1) = \({1\o {\bPsi_1^{(n)}}} \bD^{-1} {\bPsi_1^{(n)}}\)
\!\bcL(n)\! \({1\o {\bPsi_1^{(n)}}} \bD {\bPsi_1^{(n)}}\)  \;\; ,\;\;
\bcL(n-1) = \(\bP_1^{(n)} \bD {1\o {\bP_1^{(n)}}}\)
\!\bcL(n)\! \(\bP_1^{(n)} \bD^{-1} {1\o {\bP_1^{(n)}}}\)
\lab{DB-bar-L}
\ee
\be
\bP_j^{(n-1)} = \bP_1^{(n)} \bpa \(\frac{\bP_{j+1}^{(n)}}{\bP_1^{(n)}}\) 
\;,\; j \geq 1 \; ;\;\;
\bPsi_1^{(n-1)} = {1\o \bP_1^{(n)}} \;\;, \;\;
\bPsi_l^{(n-1)} = - {1\o \bP_1^{(n)}} \bpai
\(\bP_1^{(n)} \bPsi_{l-1}^{(n)}\) \;, \; l \geq 2
\lab{bar-DB-eigenf} 
\ee
\be
{\bar F}^{(n-1)} = \bP_1^{(n)} \bpa \(\frac{{\bar F}^{(n)}}{\bP_1^{(n)}}\)
\quad ,\quad
{\bar F}^{\ast\, (n-1)} = 
- \frac{1}{\bP_1^{(n)}} \bpai \(\bP_1^{(n)} {\bar F}^{\ast\, (n)}\)
\lab{bar-DB-F}
\ee
\be
\bP_1^{(n+1)} = {1\o \bPsi_1^{(n)}} \;\; ,\;\;
\bP_l^{(n+1)} = {1\o \bPsi_1^{(n)}} \bpai
\(\bPsi_1^{(n)} \bP_{l-1}^{(n)}\) \; , \; l \geq 2 ; \;\;
\bPsi_j^{(n+1)} = - \bPsi_1^{(n)} \bpa
\(\frac{\bPsi_{j+1}^{(n)}}{\bPsi_1^{(n)}}\) \;,\; j \geq 1
\lab{bar-adjDB-eigenf} 
\ee
\be
{\bar F}^{(n+1)} = \frac{1}{\Psi_1^{(n)}} \bpai \(\Psi_1^{(n)} {\bar F}^{(n)}\)
\quad,\quad
{\bar F}^{\ast\, (n+1)} = 
- \Psi_1^{(n)} \bpa \(\frac{{\bar F}^{\ast\, (n)}}{\Psi_1^{(n)}}\)
\lab{bar-adjDB-F}
\ee
where ${\bar F},{\bar F}^\ast$ are generic (adjoint) EF's of $\bcL$.
For each DB ``site'' $(n)$ the corresponding Lax operators and (adjoint)
EF's from \rf{DB-L}--\rf{bar-adjDB-F} 
define a double-KP system as in \rf{Lax-W-a}--\rf{M-generic-ab}.
Therefore, the DB orbit \rf{DB-L}--\rf{bar-adjDB-F} defines an orbit of DB
transformations for the associated two-component (matrix) KP hierarchy.

One can prove a similar statement also for generic 
DB transformations of $K\! P_1$ hierarchy:
\br
{\wti \cL} = \( F\, D\, F^{-1}\)\,\cL\,\( F\, D^{-1} F^{-1}\)    \quad ,\quad 
{\wti \P}_j = F \pa\(\frac{\P_j}{F}\)    \quad ,\quad
{\wti \Psi}_j = - {1\o F} \pai\( F\,\Psi_j\)
\lab{gen-DB} \\
{\widehat \cL} = \( F^{\ast\, -1} D^{-1} F^\ast\)
\,{\widehat \cL}\,\( F^{\ast\, -1} D\, F^\ast\)
\quad ,\quad
{\widehat \P}_j = {1\o {F^\ast}} \pai\( F^\ast \P_j\) \quad ,\quad
{\widehat \Psi}_j = - F^\ast \pa\(\frac{\Psi_j}{F^\ast}\)
\lab{gen-adjDB}
\er
where $F,F^\ast$ are generic (adjoint) EF's of $\cL$, namely,
the ``ghost'' symmetries \rf{ghost-s} commute with generic DB transformations
\rf{gen-DB}--\rf{gen-adjDB}.
\mskp
\mark
Comparing \rf{DB-L}--\rf{DB-F} with 
\rf{DB-bar-L},\rf{bar-adjDB-eigenf}--\rf{bar-adjDB-F}, and 
\rf{adjDB-L}--\rf{adjDB-F} with \rf{DB-bar-L}--\rf{bar-DB-F}, we find that
DB transformations of $K\! P_1$ hierarchy w.r.t. $\P_1$ corresponds to
adjoint-DB transformations of $K\! P_2$ hierarchy w.r.t. $\bPsi_1$, and
{\sl vice versa}.
\mskp

Let us now go back to the general $N$-component KP hierarchy.
Picking up any pair $K\! P_\a$ and $K\! P_\b$ (with $\a\!\neq\!\b$)
of one-component KP subsystems of multiple-KP hierarchy
\rf{Lax-eqs-ab}--\rf{M-generic-ab} and \rf{M-s-ag},
we can construct a DB orbit w.r.t. $\sP{\a\b}{1}$ for this pair as in
\rf{DB-L}--\rf{DB-F},\rf{bar-DB-eigenf},\rf{bar-adjDB-eigenf}--\rf{bar-adjDB-F}
by literally repeating the above contruction with the identifications:
$\cL\!\equiv\!\st{\a}{\cL},\, \bcL\!\equiv\!\st{\b}{\cL}$,
$\P_j\!\equiv\!\sP{\a\b}{j},\, \Psi_j\!\equiv\!\sPsi{\a\b}{j}$,
$F\!\equiv\!\sP{\a\g}{k},\, F^\ast\!\equiv\! \sPsi{\a\g}{k}$,
$\bP_j\!\equiv\!\sP{\b\a}{j},\, \bPsi_j\!\equiv\!\sPsi{\b\a}{j}$,
${\bar F}\!\equiv\!\sP{\b\g}{k},\, {\bar F}^\ast\!\equiv\!\sPsi{\b\g}{k}$,
where $\g\!\neq\!\a\!\neq\!\b$. Hence such DB orbit, which we will call
${\rm DB}_{(\a\b)}$ orbit, preserves the whole multiple-KP hierarchy. Also,
according to the last Remark in Section 3, ${\rm DB}_{(\a\b)}$ orbit is
equivalent to adjoint-${\rm DB}_{(\b\a)}$ orbit, {\sl i.e.}, the DB orbit
w.r.t. $\sPsi{\b\a}{1}$. Then, a general DB orbit
preserving multiple-KP hierarchy, or equivalently, $N$-component matrix KP
hierarchy, is obtained by combining the ${\rm DB}_{(\a\b)}$ orbits
for all pairs $(\a,\b)$ with $\a\!\neq\!\b$. 

We are particularly interested in DB orbits passing through the ``free''
$N$-component KP hierarchy, {\sl i.e.}, with $\st{\a}{\cL}\!=\!\st{\a}{D}$ and
hence $\t\!=\!const$ in \rf{Lax-W-a}. 
Then one can easily show that the general DB orbit for the
$N$-component KP hierarchy ${\rm DB}_{(12,13,\ldots,1N)}$ passing through the 
``free'' one is built up from a union of ${\rm DB}_{(1\b)}$ orbits of the form 
\rf{DB-eigenf}--\rf{DB-F},\rf{bar-adjDB-eigenf}--\rf{bar-adjDB-F}
with $\b\!=\! 2,\ldots,N$. Accordingly, it is
labelled by the set of integers ~$\( n_{12},n_{13},\ldots ,n_{1N};m\)$
indicating $n_{12}$ DB iterations w.r.t. $\sP{12}{1}$ or $|n_{12}|$ adjoint-DB
iterations w.r.t. $\sPsi{12}{1}$ for $n_{12}\!<\! 0$, $n_{13}$ DB iterations
w.r.t. $\sP{13}{1}$ or $|n_{13}|$ adjoint-DB iterations w.r.t. $\sPsi{13}{1}$
for $n_{13}\!<\! 0$, etc., whereas the last integer indicates $m$ steps of DB
transformations w.r.t. generic EF's $\st{1}{F}_1,\ldots,\st{1}{F}_m$ of
$K\! P_1$ subsystem, {\sl i.e.}, such that they do not belong to any of the
``ghost'' symmetry generating sets of $\sP{1\b}{k}$.
\mskp
{\bf 5. \DB Solutions}

Taking into account relations \rf{tau-ab-def} and the structure of the 
${\rm DB}_{(1\b)}$ suborbits ($\b\!=\! 2,\ldots,N$) of the form 
\rf{DB-eigenf}--\rf{DB-F},\rf{bar-adjDB-eigenf}--\rf{bar-adjDB-F} allows us
to express all non-diagonal tau-functions of the $N$-component KP hierarchy
on any site $\( n_{12},n_{13},\ldots ,n_{1N};m\)$ of the full orbit
${\rm DB}_{(12,13,\ldots,1N)}$ in terms of DB shifts of the diagonal
tau-function as follows (from now on we will use the shortened notations
$n_{1\a} \equiv n_\a$):
\be
\t_{1\a}^{(\ldots,n_{\a},\ldots)} = \eta_\a \t^{(\ldots,n_{\a}+1,\ldots)} 
\quad ,\quad
\t_{\a 1}^{(\ldots,n_{\a},\ldots)} = - \eta_\a \t^{(\ldots,n_{\a}-1,\ldots)}
\lab{tau-1-a}
\ee
\be
\t_{\a\b}^{(\ldots,n_{\a},\ldots,n_{\b},\ldots)} = 
\eta_\a\eta_\b \t^{(\ldots,n_{\a}-1,\ldots,n_{\b}+1,\ldots)} 
\quad ,\quad
\t_{\b\a}^{(\ldots,n_{\a},\ldots,n_{\b},\ldots)} = 
\eta_\a\eta_\b \t^{(\ldots,n_{\a}+1,\ldots,n_{\b}-1,\ldots)} 
\lab{tau-a-b}
\ee
where $\eta_\a\!\equiv\! sign(n_\a)$ and
$1\!<\!\a\!<\!\b\!\leq\! N$. Since $\t\,$ is the tau-function of scalar
$K\! P_1$ sub-hierarchy, the problem of finding the complete DB solution 
of the full $N$-component KP hierarchy is reduced to applying the well-known
techniques of (adjoint) DB iterations in ordinary one-component KP hierarchy
within the Sato/tau-function formulation.
Using the above techniques and taking into account the specific form of DB
orbits \rf{DB-eigenf}--\rf{DB-F} and \rf{adjDB-eigenf}--\rf{adjDB-F},
we obtain (assuming explicitly that part of the DB iterations are adjoint-DB 
ones) the following Wronskian-type expression for the diagonal tau-function:
\br
\t^{(-n_{2},\ldots ,-n_{k},n_{k+1},\ldots ,n_{N};m)}/\t^{(0,\ldots ,0;0)} 
= (-1)^{\sum_{j=2}^k n_j -k} \times
\lab{tau-full-orbit} \\
{\wti W}\Bigl\lb \st{1,k+1}{\P}\!\!\!\!{}_{1},\ldots,
\st{1,k+1}{\P}\!\!\!\!{}_{n_{k+1}},\ldots, 
\sP{1N}{1},\ldots,\st{1N}{\P}\!\!\!{}_{n_N},\st{1}{F}_1,\ldots,\st{1}{F}_m ;
\st{12}{\Psi}\!\!\!{}_{1},\ldots,
\st{12}{\Psi}\!\!\!{}_{n_{2}},\ldots, 
\sPsi{1k}{1},\ldots,\st{1k}{\P}\!\!\!{}_{n_k}\Bigr\rb
\nonu
\er
with the short-hand notation for Wronskian-type determinants:
\be
{\wti W}\llb f_1,\ldots ,f_k; f^\ast_1,\ldots ,f^\ast_l\rrb \equiv
\BDet{}{\st{1}{\pa}\!\!{}^{a-1} f_b}{\st{1}{\pa}\!\!{}^{a-1}f_{k-l+b}}{
\st{1}{\pa}\!\!\!{}^{-1}\(f_b f^\ast_c\)}{
\st{1}{\pa}\!\!\!{}^{-1}\(f_{k-l+b} f^\ast_c\)} \;\; ,\;\;
a,b\!=\! 1,\ldots ,k\!-\! l \; ,\; c\!=\! 1,\ldots ,l
\lab{Wronski-bar-def}
\ee
Recalling property \rf{SEP-EF} and the ``ghost'' symmetry flow 
Eqs.\rf{M-s-ab},\rf{M-s-generic} and \rf{N-wave-EF}, we find that we can
replace the SEP's in \rf{tau-full-orbit} (cf. \rf{Wronski-bar-def}) as
follows:
\be
\st{1}{\pa}\!\!\!{}^{-1} \Bigl(\st{1\b}{\P}\!\!\!{}_{i_\b}\sPsi{1\a}{1}\Bigr) =
\st{\a\b}{\P}\!\!\!{}_{i_\b}  \quad ,\quad
\st{1}{\pa}\!\!\!{}^{-1} 
\Bigl(\st{1\b}{\P}\!\!{}_{i_\b}\st{1\a}{\Psi}\!\!\!{}_{j_\a}\Bigr) =
\st{\a}{\pa}\!\!{}^{j_\a -1}\!\!\st{\a\b}{\P}\!\!\!{}_{i_\b} + \ldots
\lab{SEP-P}
\ee
\be
\st{1}{\pa}\!\!\!{}^{-1} \Bigl(\st{1}{F}\!\!{}_{b}\sPsi{1\a}{1}\Bigr) =
\st{\a}{F}\!\!{}_{b}   \quad ,\quad
\st{1}{\pa}\!\!\!{}^{-1} 
\Bigl(\st{1}{F}\!\!{}_{b}\st{1\a}{\Psi}\!\!{}_{j_\a}\Bigr) =
\st{\a}{\pa}\!\!\!{}^{j_\a -1}\!\!\st{\a}{F}\!\!{}_{b} + \ldots
\lab{SEP-F}
\ee
where $\a\!=\! 2,\ldots ,k$, $j_\a\!=\! 1,\ldots ,n_\a$,
$\b\!=\! k+1,\ldots ,N$, $i_\b\!=\! 1,\ldots ,n_\b$ and $b\!=\! 1,\ldots ,m$,
and where the dots in second Eqs.\rf{SEP-P}-\rf{SEP-F} indicate terms with 
lower derivatives on the corresponding EF's which cancel in the determinant.
Therefore, the Wronskian-type determinant on the r.h.s. of 
Eq.\rf{tau-full-orbit} together with \rf{SEP-P}-\rf{SEP-F} acquires the form 
of a {\em multiple} Wronskian, generalizing the double Wronskians of the
first ref.\ct{H-H}. Let us recall that all entries in the multiple Wronskian
\rf{tau-full-orbit}--\rf{SEP-F} are (derivatives of) EF's and adjoint EF's
of the respective initial $K\! P_\a$ subsystems \rf{eigenlax-a}. In the case of
``free'' initial point on the DB orbit we have:
$\spaa{\a}{s} \st{\a}{F_b} = \bigl(\spaa{\a}{1}\bigr)^s\!\!\st{\a}{F_b}$ ,
$\spaa{\a}{s} \sP{\a\b}{i} = \bigl(\spaa{\a}{1}\bigr)^s\!\!\sP{\a\b}{i}$ and
$\sP{\a\b}{i} = \bigl( -\spaa{\b}{1}\bigr)^{i-1}\!\! \sP{\a\b}{1}$.

As an example let us consider the general DB solution \rf{tau-full-orbit} in
the case $N=2$ and with a ``free'' initial $\t^{(0;0)} = 1$ :
\be
\t^{(-n;m)} = (-1)^{n-1}
\BDet{}{\pa^{a-1} F_b}{\pa^{a-1} F_{m-n+b}}{
\bpa^{c-1}{\bar F}_b}{\bpa^{c-1}{\bar F}_{m-n+b}} \;\; ,\;\;
a,b\!=\! 1,\ldots ,m-n \; ,\; c\!=\! 1,\ldots ,n
\lab{double-Wronski}
\ee
where $F_b,{\bar F}_b$ are arbitrary free EF's of $K\! P_1$ and $K\! P_2$
one-component sub-hierarchies of two-component KP hierarchy, respectively:
$\st{-}{F}_b = \int\!\! d\l\!\st{-}{\vp}_b\! (\l)\, e^{\xi (\st{-}{t},\l)}$.
The tau-functions \rf{double-Wronski}, taking into account \rf{tau-ab-def}, 
provide the following series of Wronskian solutions for DS system \rf{DS-eqs}:
\be
Q^{(-n;m)} = 4 \pa_X^2\ln\t^{(-n;m)} \quad ,\quad
\P_1^{(-n;m)} = \frac{\t^{(-n+1;m)}}{\t^{(-n;m)}} \quad ,\quad 
\Psi_1^{(-n;m)} = \frac{\t^{(-n-1;m)}}{\t^{(-n;m)}} 
\lab{DS-sol-nm}
\ee
In the particular case $m=2n$ and taking special forms of the
spectral densities of the pertinent EF's in \rf{double-Wronski}
$\st{-}{\vp}_b\! (\l)\!=\!\sum_{a=1}^{2n}\!\!\st{-}{c}\!\!\!{}_{ba}\d\(\l-\l_a\)$
with constant coefficients $\st{-}{c}\!\!\!{}_{ba}$, 
the series \rf{DS-sol-nm} with \rf{double-Wronski} contains the well-known 
$n^2$ (multi-)dromion solutions \ct{Boiti-etal} in the double-Wronskian form 
given in the first ref.\ct{H-H} (after making appropriate choice for the 
constant parameters in order to ensure reality properties).

More detailed analysis of the new series of multiple-Wronskian solutions 
\rf{tau-1-a}--\rf{SEP-F} of $N$-component KP hierarchies, in particular,
how other known soliton-type solutions are fitting there,
will be given elsewhere\foot{Let us note that
our multiple-Wronskian DB orbit differs from the DB orbit generated via
matrix EF's \ct{oevela} within Sato matrix pseudo-differential operator
formulation. The latter cannot be written compactly in a Wronskian form.}.
\mskp
{\bf 6. Conclusions}

In the present note we have shown that, given an ordinary one-component
KP hierarchy $K\! P_1$, we can always construct a $N$-component matrix KP
hierarchy, embedding the original one, in the following way. We choose $N-1$ 
infinite sets of (adjoint) EF's of the initial $K\! P_1$
(such a choice is always possible due to our spectral representation theorem
\ct{ridge}), which we use to construct an infinite-dimensional abelian algebra
of additional (``ghost'') symmetries. The one-component KP hierarchy equipped
with such additional symmetry structure turns out to be equivalent to the
standard $N$-component matrix KP hierarchy. Namely, with the help of a subset
of the ``ghost'' symmetry generating (adjoint) EF's of $K\! P_1$ we can define 
new tau-functions $\t_{\a\b}$ ($\a \neq \b \;, \; \a,\b = 1,\ldots ,N$) which,
together with $\t$ (the tau-function of the initial $K\! P_1$ hierarchy),
satisfy Hirota bilinear identites of $N$-component matrix KP hierarchy.

Furthermore, we have shown that there exists a special non-standard \DB orbit
of the initial $K\! P_1$ hierarchy, which preserves the above mentioned
additional (``ghost'') symmetry structure and which thereby generates
DB transformations of the whole $N$-component matrix KP hierarchy. This
fact allows to use the well-known DB techniques within the context of the
ordinary one-component KP subsystem $K\! P_1$ to generate soliton-like
tau-function solutions of all higher $N$-component KP hierarchies which take
the form of multiple Wronskians. In particular, we find in this way new series
of multiple-Wronskian solutions to well-known systems of integrable nonlinear 
soliton equations contained within the $N$-component matrix KP hierarchy: 
Davey-Stewartson system (for $N=2$, containing the previously obtained dromion
solutions) and $N^\pr$-wave system (for $N \geq 3$).
\mskp
{\bf Acknowledgements.} 
The authors gratefully acknowledge support by the NSF grant {\sl INT-9724747}.

{\small
\bibliographystyle{unsrt}

}
\end{document}